\documentclass[prb,twocolumn,showpacs,amsmath,superscriptaddress,psfig,amssymb]{revtex4-1}
\usepackage{amsmath}
\usepackage{amssymb}
\usepackage{graphicx}
\usepackage{dcolumn}
\usepackage{bm}
\usepackage{txfonts}
\usepackage{tabularx}
\usepackage{cases}
\usepackage[dvipdfm, pdfstartview=FitH, CJKbookmarks=true, bookmarksnumbered=true, bookmarksopen=true, colorlinks, pdfborder=001, linkcolor=blue, anchorcolor=blue, citecolor=blue]{hyperref}

\begin{document}
\title{BCS-like superconductivity in NdO$_{1-x}$F$_{x}$BiS$_{2}$ ($x$ = 0.3 and 0.5) single crystals}
\author{L. Jiao}
\affiliation{Center for Correlated Matter and Department of Physics, Zhejiang University, Hangzhou,
Zhejiang 310027, China}
\author{Z. F. Weng}
\affiliation{Center for Correlated Matter and Department of Physics, Zhejiang University, Hangzhou,
Zhejiang 310027, China}
\author{J. Z. Liu}
\affiliation{National Laboratory of Solid State Microstructures and Department of Physics, National Center of Microstructures and Quantum Manipulation, Nanjing University, Nanjing 210093, China}
\author{J. L. Zhang}
\affiliation{Center for Correlated Matter and Department of Physics, Zhejiang University, Hangzhou,
Zhejiang 310027, China}
\author{G. M. Pang}
\affiliation{Center for Correlated Matter and Department of Physics, Zhejiang University, Hangzhou,
Zhejiang 310027, China}
\author{C. Y. Guo}
\affiliation{Center for Correlated Matter and Department of Physics, Zhejiang University, Hangzhou,
Zhejiang 310027, China}
\author{F. Gao}
\affiliation{Center for Correlated Matter and Department of Physics, Zhejiang University, Hangzhou,
Zhejiang 310027, China}
\author{X. Y. Zhu}
\affiliation{National Laboratory of Solid State Microstructures and Department of Physics, National Center of Microstructures and Quantum Manipulation, Nanjing University, Nanjing 210093, China}
\author{H. H. Wen}
\affiliation{National Laboratory of Solid State Microstructures and Department of Physics, National Center of Microstructures and Quantum Manipulation, Nanjing University, Nanjing 210093, China}
\author{H. Q. Yuan}
\email{hqyuan@zju.edu.cn}
\affiliation{Center for Correlated Matter and Department of Physics, Zhejiang University, Hangzhou, Zhejiang 310027, China}
\date{\today}

\begin{abstract}
We measure the magnetic penetration depth $\Delta\lambda(T)$ for NdO$_{1-x}$F$_{x}$BiS$_{2}$ ($x$ = 0.3 and 0.5) using the tunnel diode oscillator technique. The $\Delta\lambda(T)$ shows an upturn in the low-temperature limit which is attributed to the paramagnetism of Nd ions. After subtracting the paramagnetic contributions, the penetration depth $\Delta\lambda(T)$ follows exponential-type temperature dependence at $T\ll T_c$. Both $\Delta\lambda(T)$ and the corresponding superfluid density $\rho_s(T)$ can be described by the BCS model with an energy gap of $\Delta(0)$ $\approx$ 2.0 $k_BT_c$ for both $x$ = 0.3 and 0.5, suggesting strong-coupling BCS superconductivity in the presence of localized moments for NdO$_{1-x}$F$_{x}$BiS$_{2}$.
\end{abstract}

\maketitle

The newly discovered superconductivity (SC) in Bi$_4$O$_4$S$_3$ \cite{Mizuguchi-prb,Singh} and LnO$_{1-x}$F$_x$BiS$_2$ (Ln = La, Ce, Pr, Nd and Yb)\cite{Mizuguchi-jpsj,Jha,Demura,Yazici,Xing} has attracted considerable attentions in the community. These compounds crystalize in a layered crystal structure which is composed of an alternative stacking of the BiS$_2$ double layers and the LnO blocking layer. Electron or hole doping into the LnO layers may induce SC from a semiconducting parent compound \cite{Xing}. Furthermore, recent measurements revealed a large upper critical field $\mu H_{c2}(0)$ and strong superconducting fluctuations above $T_c$ in some of these superconductors \cite{Liu,Nagao}. These features are analogous to those of high $T_c$ cuprates \cite{cuprate} and the iron-based superconductors \cite{review} and, therefore, unconventional SC with a possible high $T_c$ transition was expected in the BiS$_2$-based compounds. On the other hand, no magnetic order has been revealed in the parent compounds \cite{Mizuguchi-jpsj,Jha,Demura,Yazici}. Instead, a charge-density-wave (CDW) instability, which favors the conventional electron-phonon pairing mechanism, was theoretically discussed.\cite{Yildirim} More recently, coexistence of spin-singlet and spin-triplet paring states \cite{Yang} as well as a $g$-wave pairing state\cite{Hu} were proposed for the BiS$_2$-based superconductors; the former one arises from the strong spin-orbit coupling while the the latter one is presumably driven by electron-electron correlations. Measurements of gap symmetry may provide insights into the above controversies and are, therefore, highly desirable.

Even though intensive efforts have been devoted to the material aspects of this new superconducting family, little work has been done on their superconducting order parameters, which is partially due to the difficult growth of sizable crystals with a high sample quality. Measurements of London penetration depth and the $\mu$SR experiments on the polycrystalline Bi$_4$O$_4$S$_3$ and LaO$_{0.5}$F$_{0.5}$BiS$_2$ \cite{Lamura,Biswas,Shruti} showed evidence of an $s$-wave paring symmetry. Recently, the successful growth of single crystalline NdO$_{1-x}$F$_{x}$BiS$_{2}$ provides us a better opportunity to look into its pairing state in more detail\cite{Liu,Nagao}. Here we present measurements of the magnetic penetration depth $\lambda(T)$
for NdO$_{1-x}$F$_{x}$BiS$_{2}$ ($x$ = 0.3 and 0.5) single crystals down to 0.4 mK. The penetration depth $\lambda(T)$ and the corresponding superfluid density $\rho_s(T)$ are well described by a single-gap BCS model with strong coupling, providing compelling evidence of $s$-wave SC for NdO$_{1-x}$F$_x$BiS$_2$.

NdO$_{1-x}$F$_{x}$BiS$_{2}$ ($x$ = 0.3 and 0.5) single crystals were grown by using a flux method with
CsCl/KCl as flux.\cite{Liu} In this context, the F-concentration $x$ refers to the nominal values, which are close to the actual compositions as identified
by the energy dispersion spectrum (EDX). Precise measurements of the penetration depth changes $\Delta \lambda(T)$ were performed by utilizing a tunnel-diode oscillator (TDO) based, self-inductive technique at an operating frequencies of 7 MHz down to 0.4 K in a $^3$He cryostat, with which we can obtain a noise level as low as 0.01 PPM. The magnetic penetration depth is proportional to the shift of the resonant frequency $\Delta f(T)$, i.e., $\lambda(T) = G \Delta f(T)+\lambda(0)$,
where the $G$ factor is solely determined by the sample and coil geometries \cite{Ruslan} and $\lambda(0)$ is the penetration depth at zero temperature. The coil of the oscillator generates a tiny $ac$ magnetic field ($\mu_0H_{ac}$ $\approx$ 20 mOe), which is much smaller than the lower critical field of NdO$_{1-x}$F$_{x}$BiS$_{2}$,\cite{Jha-jap} ensuring that the measurements were performed in a Meissner state. The electrical resistivity and magnetic susceptibility were measured in a commercial Physical Properties Measurement System (PPMS) and Magnetic Properties Measurement System (MPMS), respectively.

\begin{figure}[b]
\includegraphics[width=7.5cm]{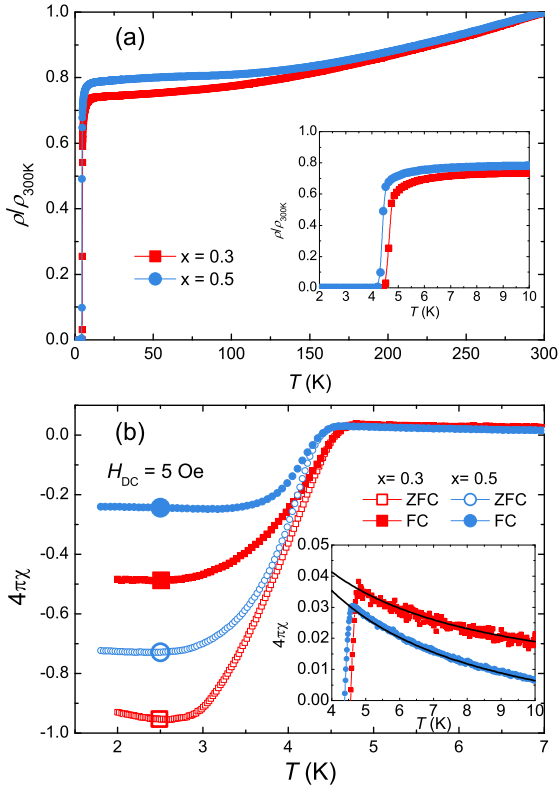}
\caption{(Color online) Temperature dependence of (a) the electrical resistivity normalized to its value at room temperature, $\rho(T)/\rho_{300K}$, and (b) the magnetic susceptibility $4\pi\chi(T)$ measured in zero-field cooling (ZFC) and field-cooling (FC) for the single crystalline NdO$_{1-x}$F$_{x}$BiS$_{2}$ ($x$ = 0.3 and 0.5). The insets expand the portions near the superconducting transition. The magnetic susceptibility in the normal state shows a weak decrease above $T_c$, which follows the Curie behavior as demonstrated by the solid lines.}
\label{fig1}
\end{figure}

In order to characterize the sample quality, we measured the temperature dependence of the electrical resistivity $\rho(T)$ and magnetic susceptibility $\chi(T)$ for NdO$_{1-x}$F$_{x}$BiS$_{2}$ ($x$ = 0.3 and 0.5), which are presented in Fig. \ref{fig1}. The electrical resistivity shows metallic behavior in the normal state for $x$ = 0.3 and $x$ = 0.5. The transition temperature $T_c$, determined from the mid-point of the sharp resistive transitions, is 4.7 K and 4.4 K for $x$ = 0.3 and 0.5, respectively. These $T_c$ values are very close to the onset transition temperature in the magnetic susceptibility $\chi(T)$. A nearly 100\% superconducting volume is observed for $x$ = 0.3 while it is reduced to 70\% for $x$ = 0.5. The NdO$_{1-x}$F$_{x}$BiS$_{2}$ system possesses a very small lower critical filed ($\mu_0H_{c1}(0) \approx 25$ Oe) \cite{Jha-jap}, thus a tiny external magnetic field may significantly broaden the superconducting transition, as shown in Fig. \ref{fig1}(b). The inset of Fig. \ref{fig1}(b) shows the magnetic susceptibility $\chi(T)$ above $T_c$, which temperature dependence can be fitted by the Curie law. i.e., $\chi(T) = C/T$. The derived Curie constants are $C$ = 0.0123 K and 0.0167 K for $x$ = 0.3 and 0.5, respectively. Consequently, an effective moment of $\mu_{eff}$ $\simeq$ 0.31 $\mu_B$ and 0.36 $\mu_B$ is estimated for $x$ = 0.3 and 0.5 respectively, where $\mu_B$ is the Bohr magneton. Such Curie behavior seems to extend to the superconducting state as evidenced by the weak upturn of the magnetic susceptibility in the low-temperature limit [see Fig. \ref{fig1}(b)] and is likely attributed to the unpaired magnetic moments of Nd$^3+$ ions.

\begin{figure}[b]
\includegraphics[width=7.5cm]{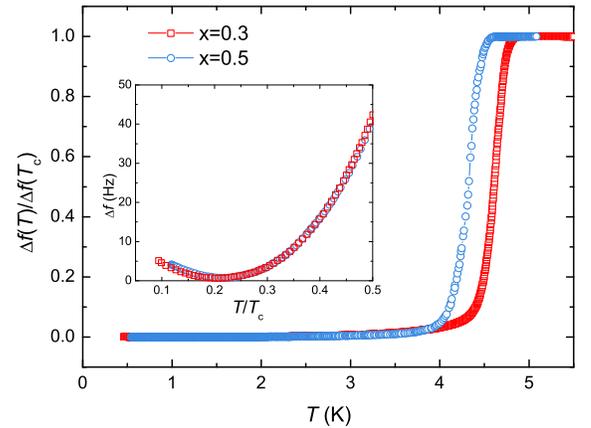}
\caption{(Color online) Temperature dependence of the frequency shift, $\Delta f(T)/\Delta f(T_c)$, normalized by its value at $T_c$ for NdO$_{1-x}$F$_{x}$BiS$_{2}$ single crystals ($x$ = 0.3 and 0.5). The inset shows the frequency shift $\Delta f(T)$ at low temperatures, which shows an upturn with decreasing temperature.}
\label{fig2}
\end{figure}

Measurements of the London penetration depth based on the TDO technique provide a unique method to probe the low-temperature excitations without an interference of magnetic field as typically encountered in the $\mu$SR and NMR experiments. In Fig. \ref{fig2}, we plot the temperature dependence of the resonant frequency shift $\Delta f(T)$ for $x$ = 0.3 and 0.5 with the $ac$ field generated along the $c$-axis; the inset expands the low-temperature part. In comparison with the magnetic susceptibility data, as shown in Fig. \ref{fig1}, a sharper superconducting transition with $T_c \simeq$ 4.5 K ($x$ = 0.3) and 4.2 K ($x$ = 0.5) is observed in $\Delta f(T)$. The so-derived $T_c$s are close to the corresponding resistive values, and the sharp transition might be related to the absence of magnetic field in the TDO-based measurements. Therefore, the TDO-based measurements can provide a precise determination of $T_c$ in those superconductors with a tiny $\mu_0H_{c1}(0)$. The consistency of the bulk and resistive $T_c$s suggest a good homogeneity of our samples.

From the inset of Fig.~\ref{fig2}, one can see that $\Delta f(T)$ for $x$ = 0.3 and 0.5 shows a pronounced upturn as temperature goes to zero. Similar behavior
was previously observed in Nd$_{2-x}$Ce$_x$CuO$_{4-\delta}$ \cite{Kokales,Prozorov00} and NdFeAsO$_{0.9}$F$_{0.1}$ \cite{Martin}, which was attributed to the paramagnetic contributions of the Nd$^{3+}$ ions. For superconductors with a significant paramagnetic background, the magnetic susceptibility in the Meissner state can be written as \cite{Martin}: 4$\pi$$\chi(T)$ = [$\sqrt{\mu(T)}\lambda_L(T)/R$]$\times$tanh[$\sqrt{\mu(T)}R/\lambda_L(T)$]-1, where $\lambda_L(T)$ is the London penetration depth, $\mu(T)$ is the normal-state paramagnetic permeability, and $R$ is a characteristic sample dimension. In this case, screening of magnetic field is described by the London equation with an effective penetration depth $\lambda(T,\mu)=\lambda_L(T)\sqrt{\mu (T)}$, and the frequency shift $\Delta f(T)$ is proportional to $\Delta\lambda(T, \mu)$ rather than the $\Delta\lambda_L(T)$, i.e., \cite{Prozorov00}
\begin{equation}
\Delta\lambda(T,\mu) = \Delta \lambda_{L}(T)\sqrt{\mu(T)} = G\cdot \Delta f(T).
\label{eq1}
\end{equation}

In this context, we assume that the magnetic susceptibility of Nd$^{3+}$ ions follows the same Curie law as in the normal state, i.e., $\chi(T)$ = $\mu(T)$-1 = $C/T$, which can reasonably describe the upturn of the frequency shift $\Delta f(T)$ in the low temperature limit (see the inset of Fig. \ref{fig2}). For superconductors with an isotropic energy gap ($s$-wave) or a nodal gap structure, the effective penetration depth $\lambda(T)$ in the presence of paramagnetic contributions takes the following expressions at $T\ll T_c$:\cite{Martin}
\begin{equation}
\lambda(T)=\sqrt{\mu(T)}\lambda(0)[1+\sqrt{\frac{\pi\Delta(0)}{2k_BT}}\textrm{exp}(-\frac{\Delta(0)}{k_BT})] ,
\label{eq2}
\end{equation}
or
\begin{equation}
\lambda(T)=\sqrt{\mu(T)}\lambda(0)[1+AT^n],
\label{eq3}
\end{equation}
where $n$ = 1 and 2 correspond to the cases of line nodes and point nodes in the superconducting energy gap, respectively.
\begin{figure}[b]
\includegraphics[width=7.5cm]{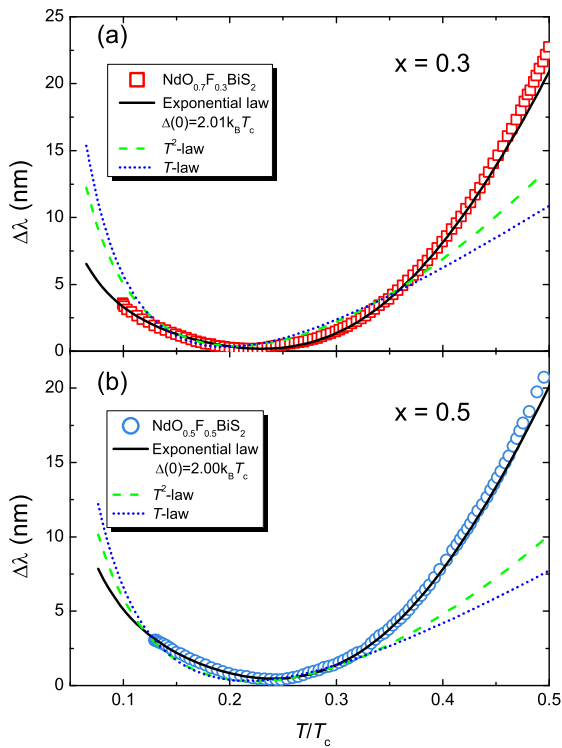}
\caption{(Color online) The change of effective penetration depth $\Delta\lambda(T)$ at low temperatures for NdO$_{1-x}$F$_{x}$BiS$_{2}$: (a) $x$ = 0.3 (a); (b) $x=$0.5. The solid lines are fits to Eq.~\ref{eq1}, while the dashed- and dotted- lines represent fits to Eq.~\ref{eq2} with $n$=1 and 2, respectively.}
\label{fig3}
\end{figure}

In Fig. \ref{fig3}, the symbols plot the temperature dependence of the effective penetration depth $\Delta\lambda(T)$ for NdO$_{1-x}$F$_{x}$BiS$_{2}$, which are converted from the frequency shift $\Delta f(T)$ with $G$ = 7.4 $\AA$/Hz and 7.8 $\AA$/Hz for $x$ = 0.3 and 0.5, respectively. The solid, dashed and dotted lines represent the fits of nodeless BCS model (Eq. 2) and nodal gaps with line ($n$ = 1) and point nodes ($n$ = 2), respectively. By taking $\lambda(0)$ as a free parameter, the fits of our experimental data to Eq. \ref{eq2} give $\lambda$(0) = 464 nm and 436 nm for $x$ = 0.3 and 0.5, which are comparable to that of LaO$_{0.5}$F$_{0.5}$BiS$_2$ [$\lambda$(0) = 484 nm] obtained from the $\mu$SR measurements \cite{Lamura}. From Fig. \ref{fig3}, one can see that the quadratic ($n$ = 2) and linear ($n=1$) temperature dependence of the London penetration depth (Eq. \ref{eq3}) fail to describe the experimental data, thus excluding the possibility of nodal SC. Instead, the BCS model (Eq. \ref{eq2}) fits nicely to the experimental data of both $x$ = 0.3 and 0.5; the fitting parameters are summarized in Table.~\ref{tab1}. The derived superconducting gaps are $\Delta(0)$ = 2.01 $k_BT_c$ and 2.00 $k_BT_c$ for $x$ = 0.3 and 0.5, respectively. These values are close to that of Bi$_4$O$_4$S$_3$ and LaOFBiS$_2$ obtained from the \textrm{$\mu$}SR experiments\cite{Biswas}, but significantly larger than that in the weak-coupling limit ($\Delta(0) = 1.76 k_BT_c$), indicating strong coupling SC in NdO$_{1-x}$F$_{x}$BiS$_{2}$.

\begin{figure}[b]
\includegraphics[width=7.5cm]{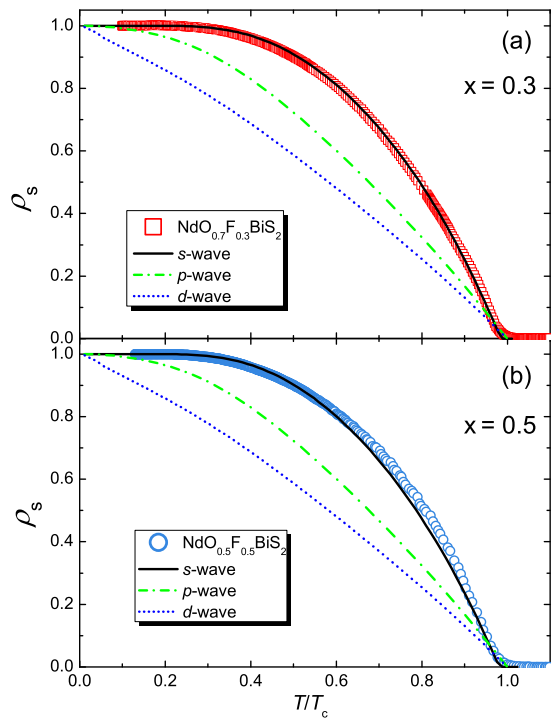}
\caption{(Color online) The normalized superfluid density $\rho_{s}(T)$ for NdO$_{1-x}$F$_{x}$BiS$_{2}$: (a) $x=$ 0.3  and (b) $x=$ 0.5 (b). The experimental data are fitted by various gap functions, i.e., $s$-wave SC (solid line), $p$-wave (dash-dotted line) and $d$-wave (dotted line).}
\label{fig4}
\end{figure}

To further analyze the gap symmetry, we derive the superfluid density $\rho_s(T)$ from the corresponding London penetration depth via $\rho_s(T)$ = $\lambda(0)^2$/$\lambda_L(T)^2$ (see Fig. \ref{fig4}), where $\lambda_L(T)$ is obtained after subtracting the paramagnetic contributions. In general, the normalized superfluid density can be calculated by:\cite{Ruslan}
\begin{equation}
\rho_s(T)=1+2\langle\int^{\infty}_{0}\frac{\partial f}{\partial E}
\frac{E}{\sqrt{E^{2}-\Delta_k^{2}(T)}}dE\rangle_{\textrm{FS}},
\end{equation}
where $f=(e^{\sqrt{{E}^2+\Delta_k^2(T)}/k_BT}+1)^{-1}$ is the Fermi
distribution function and $\langle\ldots\rangle_{\textrm{FS}}$
denotes the average over the Fermi surface. As an approximation, we assume that the energy gap $\Delta(T)$ follows the BCS-type temperature dependence:\cite{Ruslan}
\begin{equation}
\begin{aligned}
\Delta(T)=\Delta(0)\tanh\{1.82[1.018(\frac{T_c}{T}-1)^{0.51}]\}\Delta_k,
\label{eq4}
\end{aligned}
\end{equation}
where $\Delta(0)$ is the energy gap at zero temperature. Given a gap function $\Delta_k$ [= $\Delta$($\theta$, $\varphi$)] where $\theta$ and $\varphi$ denote
the polar angle and azimuthal angle, one can calculate the superfluid density $\rho_s(T)$ and fit it to the experimental data. In Fig. \ref{fig4}, three different gap functions are used to fit the experimental data $\rho_s(T)$, i.e., (i) $s$-wave: $\Delta(\theta,\varphi)$ = 1; (ii) $d$-wave: $\Delta(\theta,\varphi) = \textrm{cos}(2\varphi)$ and (iii) $p$-wave: $\Delta(\theta,\varphi) = |\textrm{sin}(\theta)|$. One can see that the experimental data $\rho_s(T)$ can be well described by the BCS model but apparently deviate from the fits of the other two models with nodes in the gap structure, providing further evidence of BCS-like superconductivity for NdO$_{1-x}$F$_{x}$BiS$_{2}$. The derived superconducting energy gaps $\Delta(0)$ are listed in Table~\ref{tab1}, which are compatible with those derived in the preceding analysis of the magnetic penetration depth $\lambda(T)$. It is noted that, for $x$ = 0.5, the fit of BCS model shows a small deviation from the experimental data $\rho_s(T)$ for $T > 0.6T_c$. At this stage, we cannot exclude the possibility of anisotropic or multiband superconductivity for $x$ = 0.5 which may fit better to the experimental data (with more fitting parameters). On the other hand, such a small deviation might be caused by some uncertainties of the derived $\lambda_0$ and the $G$-factor too.

\begin{table}[tpb]
   \caption{The superconducting parameters of NdO$_{1-x}$F$_{x}$BiS$_{2}$.}
   \label{tab1}
   \centering
   \tabcolsep=6pt
\begin{tabular}{*8c}\hline\hline
\textit{x} & $T_c$(K) & $\lambda_L(0)$ & $\Delta(0)$ by $\lambda$ & $\Delta(0)$ by $\rho_{s}$ & $C$ & $\mu_{eff}$ \\
 & K & nm & $k_BT_c$ & $k_BT_c$ & K & $\mu_B$ \\ \hline
0.3 & 4.5 & 464 & 2.01 & 2.20 & 0.0123 & 0.31 \\
0.5 & 4.2 & 436 & 2.00 & 2.15 & 0.0167 & 0.37 \\
\hline\hline
\end{tabular}
\end{table}

The above analyses of the magnetic penetration depth $\Delta\lambda(T)$ and the corresponding superfluid density $\rho_s(T)$ have shown BCS-type superconductivity for NdO$_{1-x}$F$_x$BiS$_{2}$ ($x=$0.3 and 0.5). In a conventional $s$-wave superconductor, $T_c$ can be rapidly suppressed by magnetic impurities \cite{AG}. Thus, the appearance of a BCS-like SC in the presence of localized magnetic moments of Nd ions is unusual. Indeed, evidence of nodal gap superconductivity was obtained for the Nd-based superconductors Nd$_{2-x}$Ce$_x$CuO$_{4-\delta}$ \cite{Kokales,Prozorov00} and NdFeAsO$_{0.9}$F$_{0.1}$ \cite{Martin}. Recently, it was theoretically proposed that, as a result of strong spin-orbit coupling, a dominant spin-triplet state may coexist with a spin-singlet component in the BiS$_2$-based superconductors, resulting in a crossover from nodeless SC to nodal SC with increasing the electron filling above a so-called Lifshitz point\cite{Yang}. If such a theoretical scenario is valid, then our results suggest that the filling level of NdO$_{0.5}$F$_{0.5}$BiS$_2$ is still below the Lifshitz point, and nodal SC would be expected with further increasing the F-doping concentration, which might push the Fermi level close to a van Hove singularity. Indeed, the recent ARPES experiments revealed that the deficiency of Bi element in NdO$_{1-x}$F$_x$BiS$_{2}$ may result in a reduction of the actual filling level,\cite{Ye,Zeng} meaning that the samples with $x$ = 0.5 are likely below the Lifshitz point. Therefore, it is highly desired to systematically study the samples with higher doping concentrations. Unfortunately, the growth of single crystals with rich F-content has not been successful yet and further efforts are badly needed.

In summary, we have performed measurements of magnetic penetration depth $\Delta\lambda$(T) for the NdO$_{1-x}$F$_{x}$BiS$_{2}$ ($x$ = 0.3 and 0.5) single crystals by utilizing a TDO-based technique down to 0.4 K. Strong evidence of BCS-like SC with a large gap size is found in both the low-temperature penetration depth $\lambda(T)$ and its corresponding superfluid density $\rho_s(T)$. A Curie-like paramagentic contribution of Nb$^{3+}$ shows up in the superconducting state which is rare in a BCS-like superconductor. Future works are needed to clarify the gap structure in the overdoped region and, therefore, to test the currently available theories.

We are grateful to Q. H. Wang and C. Cao for helpful discussions. This work was supported by the National
Basic Research Program of China (No. 2011CBA00103), the National Nature Science Foundation of China (No. 11174245) and the Fundamental Research Funds for the Central Universities.

\end{document}